%% file: pnaa.tex
\documentstyle[epsf]{l-aa}

\input abb.tex

\begin{document}
\thesaurus{03(16.02.1; 
              16.02.1; 
              07.19.1: 
                      }
\title 
{Two New Planetary Nebulae Discovered in a Galaxy Search in the Southern
Milky Way}

\author{R.C.~Kraan-Korteweg\inst{1,4} \and A.P.~Fairall\inst{2}
\and P.A. Woudt\inst{2,4} \and G.C. Van de Steene\inst{3,4}}
              
\offprints {Ren\'ee C. Kraan-Korteweg}

\institute{Observatoire de Paris-Meudon, D.A.E.C., 5 Place Jules Janssen,
92195 Meudon Cedex, France
\and
Department of Astronomy, University of Cape Town, 
Rondebosch, 7700 South Africa
\and
ESO, Casilla 19001, Santiago 19, Chile
\and
formerly Kapteyn Astronomical Institute, Postbus 800, Landleven 9,
9700 AV Groningen, The Netherlands}

\date{Received date;accepted date}
\maketitle
\markboth
{R.C.~Kraan-Korteweg et~al.: Observations of PN-Candidates}
{Kraan-Korteweg et al.}

\begin{abstract}

Spectroscopic observations have been carried out for eleven
objects believed to be planetary nebulae on the basis of 
their optical appearance.
They were discovered in an ongoing deep search for galaxies
in the Southern Milky Way (Kraan-Korteweg \& Woudt 1994).  The objects
were observed with the 1.9m telescope of the South African Astronomical
Observatory during our program for obtaining redshifts of obscured 
galaxies in the ``Zone of Avoidance''.  
Of the eleven objects, three proved too faint for a definite classification, 
four were galaxies with radial velocities between v=3920 km/s and 
v=14758 km/s, but four were confirmed as planetary nebulae (PNE).  
Their relative line strengths and radial velocities have been determined.  
The PNE are on average fairly large ($23\arcsec-30\arcsec$).  
Two of them (PNG 298.3+06.7 and PNG 323.6-04.5) were previously unknown;  
for these we show H$\alpha$ and [O\,{\sc iii}] images. 
\end{abstract}


\section{Introduction}
The foreground absorption of our Galaxy obscures about 25\% of the 
extragalactic sky. In order to narrow this ``Zone of Avoidance'', we 
have been searching the film copies of the ESO/SERC IIIaJ survey in the
southern Milky Way for partially obscured galaxies. Using a 
50 x magnifying viewer, we have identified all galaxies
and galaxy candidates to a lower diameter limit of D=$0\farcm2$,
and $-$ depending on their surface brightness $-$ to a magnitude limit of
about $B_J \approx 19\fm0 $-$ 19\fm5$. To date, over 10,000 previously
uncatalogued galaxies have been identified in the area $266\deg \la \ell \la 
340\deg, |b| \la \pm 10\deg$. For further details 
the reader is referred to Kraan-Korteweg, 1989, Kraan-Korteweg \& Woudt 1994,
and references therein.

In the course of this deep search we have come across a number of 
extended objects having sharp edges that suggested their being PNE 
rather than external galaxies.  
In section 2, the optical properties of these 11 prospective
PNE are described.  These objects were observed spectroscopically with
the 1.9m telescope of the SAAO during our redshift observations of
galaxies in the Zone of Avoidance (Kraan-Korteweg et al. 1994, 1995,
Kraan-Korteweg, Fairall \& Balkowksi, 1995).  The spectroscopic
observations are described in section 3.  Four of the eleven objects are
confirmed as PNE.  The resulting spectra are displayed and their
relative line strengths discussed.  Two PNE were previously unknown.  In
section 4, the [O\,{\sc iii}] and H$\alpha$ images obtained at the 90cm
Dutch telescope of the European Southern Observatory (ESO) of these PNE
are displayed, followed by a discussion of the two new PNE in the
last section. 

\section{The planetary nebula candidates}
The 11 prospective PNE are shown in Figure 1. 
\begin{figure*}[t]
\hfil\epsfxsize 18cm \epsfbox{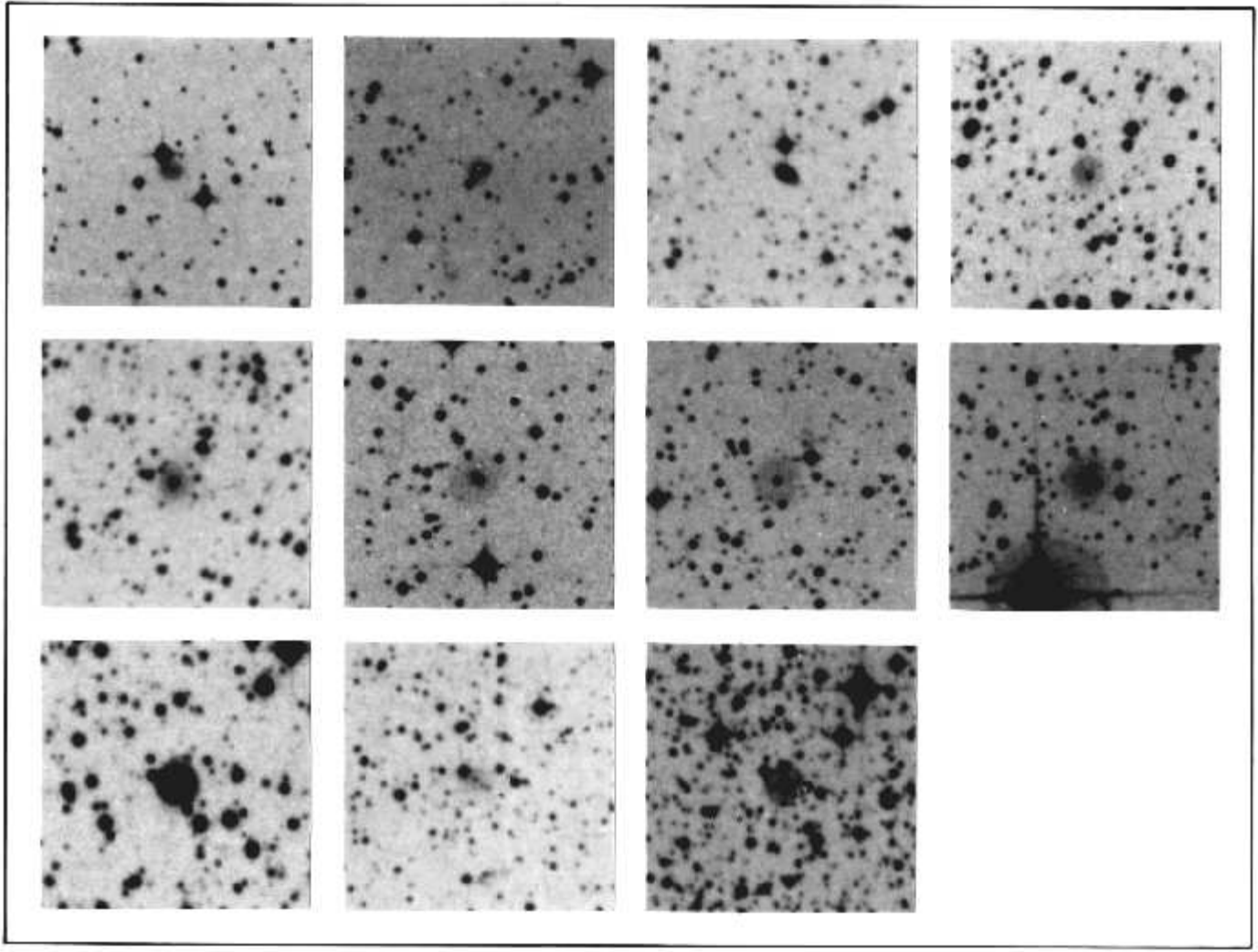}\hfil
\caption{Images of the prospective PN-candidates as seen on the film 
copies of the ESO/SERC IIIaJ survey during the galaxy search. The images are
enlarged by a factor of 20 and are $2\farcm2$ x $2\farcm2$. North is
up, east is left. The images are ordered from left to right, 
top to bottom according to the sequence in Table 1.}
\label{f1}
\end{figure*}  

As mentioned above, they were classified as PN-candidates based on their
optical appearance on the film copies of the ESO/SERC IIIaJ sky survey. 
These extended objects were not believed to be galaxies because of their
sharp edges $-$ too sharp for the characteristic morphology of external
galaxies.  Furthermore half of the objects reveal a distinct central
core.  Two of the objects (3 and 9) display a high surface brightness. 

The positions and other parameters of the objects 
are listed in Table~1. The entries in Table~1 are as follows:
\begin{table*}[t]
\caption{PN-candidates observed spectroscopically at the SAAO}
{\scriptsize
\input table1.tex
}
\label{t1}
\end{table*}

\noindent {\it Column 1:} Running number, corresponding to the images
displayed in Figure~1.

\noindent {\it Column 2:} Identification of the galaxy.  E denotes the
identification in the ESO-Uppsala Survey (Lauberts, 1982), RKK refers to
the catalogue of galaxies in the southern Milky Way in the Hydra/Antlia
extension (Kraan-Korteweg 1996, henceforth RKK96), and WKK to its
extension towards the Great Attractor region (Woudt and Kraan-Korteweg,
1996 [WKK96]).  RKK-P1 refers to a separate listing of objects
identified in RKK96 which most likely are PNE, not galaxies.  The
asterix indicates that the object has an entry in the IRAS PSC (cf. 
Table 2). 

\noindent {\it Column 3 and 4:} Right Ascension and Declination (epoch
1950.0).  The positions were measured with the measuring machine
Optronics at the ESO in Garching and have an accuracy of about 1 arcsec. 

\noindent {\it Column 5 and 6:} Galactic longitude $\ell$ and latitude
$b$. 

\noindent {\it Column 7, 8 and 9:} The field of the ESO/SERC survey on
which the object was measured and the x and y offset in mm from the
center of that field. 

\noindent {\it Column 10:} Large and small diameter (in arcsec).  The
diameters are measured to an approximate isophote of 24.5 mag
arcsec$^{-2}$ and have a scatter of $\sigma \approx 4\arcsec$. 

\noindent {\it Column 11:} 
Apparent magnitude B$_J$.  The magnitudes are estimates from the IIIaJ
film copies of the ESO/SRC Survey based on the above given diameters and
an estimate of the average surface brightness.  A preliminary analysis
of this data for the galaxy survey finds a linear relation from the
brightest to the faintest magnitudes (B$_J \approx 19\fm5$) with a
scatter of only $\sigma \approx 0\fm5$. 

\noindent {\it Column 12:} Object status after spectroscopy 
and/or heliocentric velocity (cf. next section). 

\section{Spectroscopic Observations} The spectroscopic observations of
the 11 prospective planetary nebulae were made during observing runs in
March 1991, 1993 and April 1994 at the South African Astronomical
Observatory (SAAO) at Sutherland.  The 1.9m Radcliffe reflector with
``Unit'' spectrograph and reticon photon-counting detector was used. 
The dispersion was approximately 2.8\AA\ per pixel.  The reduction
procedures are described in detail in Kraan-Korteweg, Fairall \&
Balkowski (1995). 

Three of the 11 objects (2, 8, and 10) were too faint, their surface 
brightness too low, for a definite conclusion to be made about their
nature.  No emission was detected.  Spectroscopy of the other 8
objects proved four (1, 4, 5, and 6) to be galaxies despite their
optical appearance.  Their redshifts are given in the last column of
Table 1.  The remaining four objects (3, 7, 9 and 11) were confirmed as
PNE.  A cross-identification with the Strasbourg-ESO Catalogue of
Galactic Planetary Nebulae (Acker et al.  1992) indicates that the two
planetary nebulae WKK171-090 and WKK136-337 were previously unknown.  In
accordance with the astronomical nomenclature convention for PN by their
galactic coordinates (Acker at al.  1992), they are forthwith called PNG
298.3+06.7 and PNG 323.6-04.5

The spectra of the four PNE are illustrated in Figure 2.  The identified
emission lines are marked.  The spectral line fluxes of the PN with the
higher surface brightness on the plates, the previously known 
E166-P18 = PNG 275.5-01.3 and E131-P01 = PNG 299.5+02.5, are higher compared 
to the two previously unknown PNE.  
For the planetary PNG 323.6-04.5, the Balmer line is relatively weak.  
Some coincidence is found at the lines NeIII, [O\,{\sc iii}](4363\AA) and
HeII, but they are not believed to be real.  

\begin{figure*} 
\hfil \epsfxsize 18cm\epsfbox{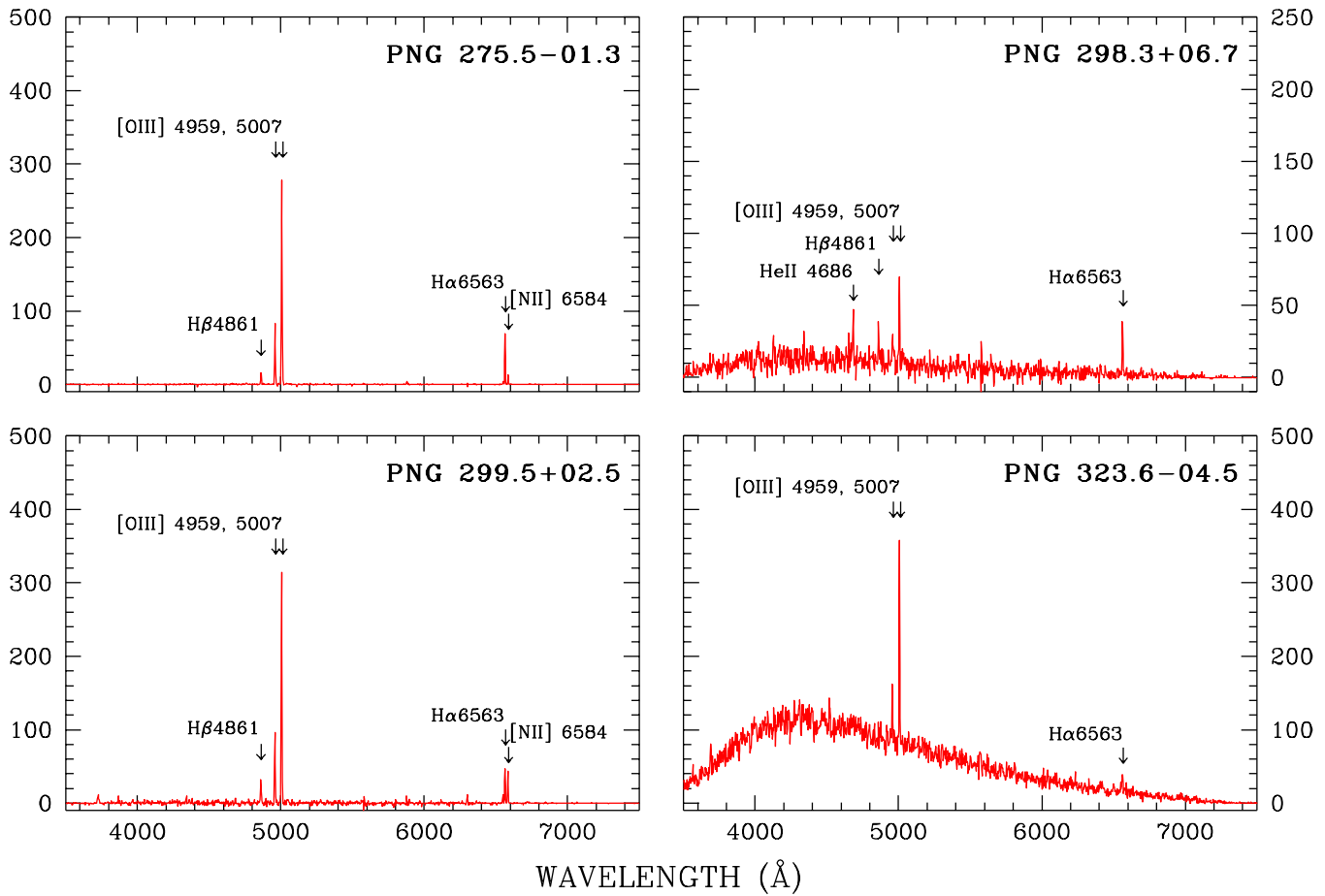}\hfil 
\caption{Spectra of the four confirmed PNE obtained with the 1.9m 
telescope of the SAAO.}
\label{f2} 
\end{figure*}

Of the 4 spectroscopically observed PNE, only PNG 275.5-01.3 and PNG 
323.6-04.5 
are listed in the IRAS PSC. Their respective fluxes are given in Table 2.
\begin{table}[h]
\caption{Planetary Nebulae in the IRAS PSC}
{\tabcolsep0.5mm
{\scriptsize
\input table2.tex
}}
\label{t2}
\end{table}

\subsection{Relative line strengths and radial velocities of the 4 planetary
nebulae}

The use of narrow slits and the general characteristics of the
spectrograph do not permit absolute calibration of the intensity scale. 
The observation of spectrophotometric standard stars has enabled us to
measure the relative instrumental response and the spectra have been
suitably corrected.  Relative instrumental calibration was done by
observing a spectrophotometric standard star -- in this case LTT 7379. 
PNG 275.5-01.3 was observed in March 1991.  Unfortunately, no
spectrophotometric standard star was observed at that time due to
adverse weather conditions and we therefore have no
accurate way to bridge the H$\alpha$ region response to the H$\beta$
region response for this planetary. A rough estimate was made from 
the other observing runs, with sufficient accuracy to establish that 
there is a steep Balmer decrement, and thus a large amount of reddening. 

The ``Unit'' Spectrograph with Reticon Photon Counting System is
generally used for radial velocity work and is not an ideal instrument
for measuring line strengths.  However, we observed the
planetary nebula NGC 3132 = PNG 272.1+12.3 at RA=$10^h04^m55^s$, 
Dec=$-40\deg11\arcmin29\arcsec$ with known line strengths as a standard.
This PN  has a similar diameter compared to the PNE in our sample 
(D=30$\arcsec$).  The results for the 4 PNE and our ``standard PN'' 
are summarized in Table 3. 
\begin{table*}[t] \caption{Emission line strengths for the four
confirmed PN and the calibrator NGC 3132=PNG 272.1+12.3} 
{\tabcolsep0.5mm 
{\scriptsize 
\input table3.tex

}} 
\label{t3}
\end{table*}

The header line lists the identified emission lines with their
respective wavelengths in {\AA}ngstr{\o}m.  The first column lists the
name of the PNE and source: for the 2 previously known PNE and the
standard, the data as given in the Strasbourg-ESO catalogue of Galactic
Planetary Nebulae (Acker et al. 1992) are listed (identified by
``Acker'' in the first column) next to our results (identified with
``SAAO'').  This allows an assessment of the accuracy of the
observational procedures for extracting relative line strenghts. 
Blow-up plots were used to measure the area under the lines and
establish the linestrengths.  In conformance to the Acker et al. 
catalogue, the fluxes are expressed relative to an H$\beta$-line
intensity of 100.  The second to last column lists the logarithm of the
absolute flux of the H$\beta$ line (in mW m$^{-2}$) for the standard PN
and the 2 previously known PNE.  Radial velocities were derived from the
stronger emission lines.  The centroids of the emission lines
were again determined from the blow-up plots.  The resulting values are
entered in a second line in Table 2 at the respective wavelengths.  The
last columns lists the mean radial velocity of the 4 PNE. 

\section{CCD-Imaging of the 2 new planetary nebulae PNG 298.3+06.7 
and PNG 323.6-04.5}

Shortly after the spectroscopic identification of the 2 new PNE, 
CCD images were taken of PNG 298.3+06.7 and PNG 323.6-04.5 in both 
H$\alpha$ and [O\,{\sc iii}] using the Dutch 0.9m telescope of the European 
Southern Observatory. This telescope is equipped with a TEK512 CCD
camera, giving a field of view of $3\farcm8$x$3\farcm8$ and a pixelsize of
0\farcs44/pixel.

\begin{figure*}[t]

\caption{Images of the PNE PNG 298.3+06.7 (top panels) and PNG 323.6-04.5 
(bottom panels) obtained with the 0.9m Dutch telescope
at la Silla, ESO. Left panel: H$\alpha$ (ESO filter 387), right panel: 
[O\,{\sc iii}] (ESO filter 688). Note that North is up and East is right.}
\label{f4}
\end{figure*}
The resulting CCD images are shown in figure 3.
The top panels illustrate PNG 298.3+06.7: the exposure time for the 
H$\alpha$ and the [O\,{\sc iii}] images were 1200 sec and 1800 sec 
respectively.  The basic reductions such as bias subtraction and flat 
fielding were done within the MIDAS image reduction package. 

The bottom panels show PNG 323.6-04.5. The same H$\alpha$ and [O\,{\sc iii}]
filters were used for these observation. In this case however, both images 
have an exposure time of 600 seconds. 

Due to non-photometric weather conditions these images are not flux 
calibrated. As discussed in the next section, they do, however, give
a fair impression of the optical morphology and show the similarity in 
appearance between both new PNE.

\section{Discussion of the new planetary nebulae}

These new optically discovered PNE appear very spherical.  They are of
rather faint surface brightness, which is probably why they have gone
undetected. 
 
Both new PNE are probably closer than 8 kpc: because of the concentration 
of PNE towards the galactic bulge, the distribution of angular diameters 
of PNE has a strong peak between 4 and 12 arcsecs -- although nearer 
nebulae with angular diameters up to 2 arcminutes have been observed.  
PNG 298.3+06.7 and PNG 323.6-04.5 both have an angular size of 
$20\farcs3$ and are among the larger PNE.
Most bulge planetary nebulae with a distance of typically 6 - 10 kpc
have diameters below 10$\arcsec$. Hence, the 2 newly uncovered PNE
are likely to be closer than the Galactic center distance of 8 kpc. 

Though these nebulae look very similar there is a striking difference
between them.  The central star of PNG 298.3+06.7 is only a few
times brighter compared to the mean of the nebular brightness, while
the star at the center of PNG 323.6-04.5 is more than ten times 
brighter than the nebula.  The central star of PNG 298.3+06.7 clearly
is pointlike, while the central part of PNG 323.6-04.5 might be slightly 
resolved -- but this needs to be confirmed.  The central star of 
PNG 323.6-04.5 is very bright and the limb of the nebula more defined 
than PNG 298.3+06.7. 

The dust temperatures of the PNE detected by IRAS were calculated
according to the formalism given in Van de Steene \& Pottasch (1993,
1995).  For PNG 323.6-04.5, the dust temperature is found to be 86 K.
This low dust temperature together with the very low far IR flux of 
1.1$\cdot 10^{-13}$ W m$^{-2}$ (cf. Table 2) is indicative of an 
evolved nebula with little dust. 
The observed H$\alpha$/H$\beta$ ratio, as compared to the
recombination value of 2.85 (Aller, 1984), results in a logarithmic
extinction at H$\beta$ of:\\
c(H$\beta$) = 3.096 log(H$\alpha$ / (H$\beta$ / 2.85))\\
corresponding to a visual extinction of A$_{V}$ = $2.1$
c(H$\beta$).  The extinction values determined from the Balmer
lines by Tylenda et al. (1992) from better spectra are $0.16$,
$2.2$ and $1.0$ for PNG 272.1+12.3, PNG 275.5-01.3 
and PNG 299.5+02.4 compared to $0.5$, $1.7$ and $0.5$,
the values determined from the SAAO spectra.  The logarithmic 
extinction of the new PNE are $0.5$ for PNG 298+06.7 and
$1.7$ for PNG 323-04.5, where the higher extinction for
PNG 323-04.5 is evident already in the images. 

The flux values were corrected for interstellar extinction using the
literature values and the galactic extinction law of Seaton
(1979).

The photon energy distribution determines the degree of excitation in
the nebula reflected by the excitation class (E.C.).  The excitation
class was calculated using the extinction corrected line fluxes,
according to the scheme by Dopita $\&$ Meatheringham (1991):\\
E.C. $=$ 5.54 (F$_{\lambda 4686}$/F$_{H\beta}$ $+$ 0.78) 
\hspace{0.3cm} for 5.0 $\leq$ E.C.\\
E.C. $=$ 0.45 (F$_{\lambda 5007}$/F$_{H\beta}$) 
\hspace{1.4cm} for 0.0 $\leq$ E.C. $\leq$ 5\\ 
This definition is to some extent sensitive 
to the chemical abundances of helium and of oxygen, and also on the
ionization parameter.  To estimate the stellar temperature (T$_{eff}$), 
we used the correlation between E.C. and T$_{eff}$ (Dopita $\&$
Meatheringham, 1991).  We find E.C.$=$ 4.1 and T$*=$80,000 K for 
PNG 272.1+12.3.  Preite-Martinez et al. (1989, 1991) determined an
E.C.$=$ 6 and a T$_{eff}$=99,400 K using the energy balance method.  For PNG
275.5-01.3, the excitation class determined from the [O\,{\sc iii}] line
is 5.9. This translates to T$_{eff}$$>$100,000 K. In the latter case
one would, however, expect to see He\,{\sc ii} which is not observed. 
Preite-Maritnez et al. determined E.C.$=$5 and an energy balance
T$_{eff}$$=$81,600 K.  For PNG 299.5+02.4 we determined E.C.$=$3.7
corresponding to T$_{eff}$=75,000 K, while Preite-Martinez et al. 
determined E.C.$=$5 and T$_{eff}$$=$71,800 K. 

The blue stellar continuum of the white dwarf is visible in both spectra
of the new PNE.  According to what we see in the images the stellar
continuum appears stronger for PNG 323.6-04.5 than for PNG 298.3+06.7. 

For PNG 298.3+06.7 the extinction-corrected [O\,{\sc iii}]$\lambda$5007 
line (208) is a bit weaker than H$\alpha$, indicative of a
relatively low excitation.  But the He\,{\sc ii}$\lambda$4686 line
seems exceptionally strong! This line is usually much weaker than
H$\beta$ -- even in PNE which are overabundant in helium.  It could, of 
course, be that the hydrogen lines are exceptionally weak and that 
this PN belongs to the class of hydrogen deficient PNE, but this 
would indicate a very high excitation nebula. 
For PNG 323.6-04.5 the [O\,{\sc iii}] line is very strong. It is therefore
peculiar that no He\,{\sc ii} is apparent in this high excitation nebula
unless, of course, H$\alpha$ is very weak and this is also a hydrogen
deficient nebula. Future observations will tell.
The present information is insufficient to determine either electron 
temperature or density for the new PNE.
\vspace{0.2cm}

\noindent
Our galaxy search is still ongoing and further prospective PNE have
been discovered meanwhile. We will report on those as soon as we have
follow-up observations.

\acknowledgements
{The authors would like to thank the night assistants Francois van Wijk
and Frans Marang as well as the staff at the SAAO for their hospitality. 
The help of Adwin Boogert in obtaining the images of PNG 323.6-04.5 with the 
Dutch telescope on La Silla (ESO) is greatfully acknowledged.
While in Groningen, the research by RCKK has been made possible by a 
fellowship of the Royal Netherlands Academy of Arts and Sciences. APF and
PAW are supported by the South African FRD.}

\end{document}

%% file: abb.tex
\def\la{\mathrel{\hbox{\rlap{\hbox{\lower4pt\hbox{$\sim$}}}\hbox{$<$}}}}
\def\ga{\mathrel{\hbox{\rlap{\hbox{\lower4pt\hbox{$\sim$}}}\hbox{$>$}}}}

\def\arcmin{\hbox{$^\prime$}}
\def\arcsec{\hbox{$^{\prime\prime}$}}

\def\fm{\hbox{$.\!\!^{m}$}}
\def\fs{\hbox{$.\!\!^{s}$}}

\def\farcm{\hbox{$.\mkern-4mu^\prime$}}
\def\farcs{\hbox{$.\!\!^{\prime\prime}$}}

\def\deg{{^\circ}}

%% file: table1.tex
\begin{tabular*}{18cm}{@{\extracolsep\fill}r l r r r r r r r r r l}
\noalign{\smallskip} 
\hline
\hline
\noalign{\smallskip} 
\ RN & Identification& RA\ \ \ \ \ & Dec \ \ \ \   & gal. $\ell$  & gal. $b$ &SRC& X \ \ & Y \ \    & Dxd \           & B$_J$ \  & \ \ \ \ \ Status \\ 
   &                  & 1950.0 \ \  & 1950.0 \ \    &              &          &   & (mm) & (mm)    & ($\arcsec$)\ \  & mag & \\
(1) & \ \ \ \  (2)    & (3)\ \ \ \ \  & (4)\ \ \ \ \ & (5)\ \ \    & (6)\ \   &(7)& (8) \  & (9) \ & (10)\ \         &(11) & \ \ \ \ \ \ (12)\\
\noalign{\smallskip}
\hline
\noalign{\smallskip}
\input table1.dat
\noalign{\smallskip}
\hline
\end{tabular*}

%% file: table1.dat
    1&RKK 0008  & 8$^h$31$^m$19$\fs$7&-55$\degr$55$\arcmin$40$\arcsec$&272.17& -9.58&165&-127.8& -52.8&13x13&18.2&gal. at $v=4219$km/s\\
    2&RKK-P1    & 8$^h$39$^m$10$\fs$1&-55$\degr$42$\arcmin$\ 2$\arcsec$&272.65& -8.56&165& -69.3& -38.1&16x12&18.2&too faint, no em. det.\\
    3&E166-P18*& 9$^h$29$^m$\ 8$\fs$8&-52$\degr$56$\arcmin$43$\arcsec$&275.52& -1.33&166&  65.5& 109.4&23x15&16.7&PN confirmed\\
    4&RKK 1833  & 9$^h$56$^m$58$\fs$3&-63$\degr$41$\arcmin$21$\arcsec$&285.26& -7.17& 92&-115.2&  68.4&20x20&17.5&gal. at $v=11522$km/s\\
    5&RKK 2432  &10$^h$21$^m$16$\fs$6&-65$\degr$31$\arcmin$46$\arcsec$&288.47& -7.17& 92&  27.3& -25.9&26x19&17.0&gal. at $v=14758$km/s\\
    6&RKK 2450  &10$^h$21$^m$51$\fs$4&-49$\degr$31$\arcmin$\ 7$\arcsec$&279.97&  6.41&214& -71.7&  24.2&34x28&16.4&gal. at $v= 3920$km/s\\
    7&WKK171-090&12$^h$15$^m$51$\fs$7&-55$\degr$37$\arcmin$25$\arcsec$&298.30&  6.67&171&  73.7& -33.9&28x28&16.9&{\bf new PN}\\
    8&WKK171-052&12$^h$19$^m$\ 5$\fs$0&-54$\degr$\ 6$\arcmin$24$\arcsec$&298.58&  8.24&171& 101.9&  46.3&35x26&16.9&too faint, no em. det.\\
    9&E131-P01 &12$^h$21$^m$\ 8$\fs$4&-59$\degr$56$\arcmin$34$\arcsec$&299.51&  2.47&131&-124.9&    .1&30x28&16.1&PN confirmed\\
   10&WKK172-334&12$^h$31$^m$\ 2$\fs$9&-56$\degr$22$\arcmin$10$\arcsec$&300.50&  6.14&172& -60.8& -73.4&16x15&18.2&too faint, no em. det.\\
   11&WKK136-337*&15$^h$46$^m$18$\fs$4&-59$\degr$49$\arcmin$44$\arcsec$&323.63& -4.51&136& -28.2&  10.5&26x24&17.0&{\bf new PN}\\

%% file: table2.tex
\begin{tabular*}{8.8cm}{r @{\extracolsep\fill} l c c c c c c c}
\noalign{\smallskip} 
\hline
\hline
\noalign{\smallskip} 
 RN & \ Identi- & IRAS-name & \multicolumn{4}{c} {IRAS-flux} & flux \\
    & fication  &           & f$_{12}$ & f$_{25}$& f$_{60}$ & f$_{100}$ & qual.\\
(1) & \ \ \ (2)   & (3)       & (4)      & (5)     & (6)      &(7)        & (8) \\
\noalign{\smallskip}
\hline
\noalign{\smallskip}
  3 & PNG 275.5-01.3 & 09291-5256 & 0.25L & 1.41 \ & 2.94 \ &  5.74 & -CCD \\ 
 11 & PNG 323.6-04.5 & 15463-5949 & 0.37L & 0.34 \ & 1.32 \ & 20.04L & -BB- \\
\noalign{\smallskip}
\hline
\end{tabular*}

%% file: table3.tex
\begin{tabular*}{18cm}{l @{\extracolsep\fill}*{16}{c}}
\noalign{\smallskip} 
\hline
\hline
\noalign{\smallskip} 
 PN            & [OII] & [NeIII] & H$\gamma$ & [OIII]& HeII & H$\beta$ & [OIII] & [OIII] & HeI  & [NII] & H$\alpha$ & [NII] & [SII] & [SII] & lg I(H$\beta$) & Vel.\\
               & 3727  & 3868    & 4340      & 4363  & 4686 & 4861     & 4959   & 5007   & 5876 & 6548  & 6563      & 6584  & 6717  & 6731  &             &     \\
\noalign{\smallskip}
\hline
\noalign{\smallskip}
 PNG 272.1+12.3 (Acker)&       &         &           &    4  &     5 &     100 &        &   1021 &   14 &       &       319 &  393 &    32  &  32  & -10.45 & -10$\pm$ 3 \\ 
 PNG 272.1+12.3 (SAAO) & 321   & 76      &  30       &    4  &       &     100 &    255 &    939 &      &   279 &       416 &  966 &        &      &        &            \\  
\noalign{\smallskip}
 Radial velocity&-53   &         & -48       &       &       & -32     &    -7  &    -13 &      &       &   +5      &  +28 &        &      &        & -15$\pm$40 \\
\noalign{\smallskip}
\hline
\noalign{\smallskip}
 PNG 275.5-01.3 (Acker)&     &         &           &     - &     - &     100 &        &   1844 &   42 &       &      1496 &  339 &     12 &  23  & -13.00 \\
 PNG 275.5-01.3 (SAAO) &     &         &  30       &       &       &     100 &    427 &   1568 &      &   279 &      1000:&  188:&        &      &        \\
\noalign{\smallskip}
 Radial velocity &     &         &           &       &       &     +37 &     +2 &     +2 &      &       &       +29 &      &        &      &        & +17$\pm$16 \\
\noalign{\smallskip}
\hline
\noalign{\smallskip}
 PNG 298.3+06.7 (SAAO)&     &         & 63        &       &   122 &     100 &     93 &    216 &   40 &       &       412 &   50 &        &      &        \\
\noalign{\smallskip}
 Radial velocity &     &         & -35       &       &       &     -37 &  (-6)  &    -48 &      &       &       -32 &      &        &      &       & -38$\pm$ 6 \\
\noalign{\smallskip}
\hline
\noalign{\smallskip}
 PNG 299.5+02.5 (Acker)&     &         &           &     - &     - &     100 &        &    807 &   -  &       &       610 &  486 &     83 & 82  & -12.60 & -10$\pm$12 \\
 PNG 299.5+02.5 (SAAO) &     &         &           &       &       &     100 &    271 &    897 &      &    87 &       403 &  352 &        &     &        &            \\
\noalign{\smallskip}
 Radial velocity &     &         &           &       &       &     -41 &    -10 &    -16 &      &       &        +7 &  -25 &        &     &        & -17$\pm$16 \\
\noalign{\smallskip}
\hline
\noalign{\smallskip}
 PNG 323.6-04.5 (SAAO)&     &         &           &       &       &  100$^{(a)}$ &    788 &   2538 &      &       &      1031 &  352 &        &     &        &            \\
\noalign{\smallskip}
 Radial velocity &     &         &           &       &       &      -9 &    -87 &    -38 &      &       &       -37 &      &        &     &        & -41$\pm$25$^{(b)}$ \\
\noalign{\smallskip}
\hline
\end{tabular*}

\noindent
$^{(a)}$ H$\beta$ is only marginally present and the relative line strengths should be regarded as tentative only\\
$^{(b)}$ higher weight was given to the velocity determined from 
[O~{\sc iii}]~$\lambda$5007, the stronger isolated line in
the spectrum